\documentclass[twoside]{article}
\usepackage{ShortUmodule}

\title{Unifying two Graph Decompositions with Modular Decomposition$^0$}
\author{
\normalsize{Binh-Minh Bui-Xuan}$^1$
\and
\normalsize{Michel Habib}$^2$
\and
\normalsize{Vincent Limouzy}$^2$
\and
\normalsize{Fabien de Montgolfier}$^2$
}
\begin{document}
\setcounter{footnote}{0}
\footnotetext{Research supported by the French ANR project ``Graph Decompositions and Algorithms (GRAAL)''}
\setcounter{footnote}{1}
\footnotetext{LIRMM, Univ. Montpellier \texttt{buixuan@lirmmfr}}
\setcounter{footnote}{2}
\footnotetext{LIAFA, Univ. Paris Diderot. \texttt{\{habib,limouzy,fm\}@liafa.jussieu.fr}}
\date{}
\maketitle
\vspace{-0.75cm}
\begin{abstract}
We introduces the \emph{umodules}, a generalisation of the notion of graph module.
The theory we develop captures among others undirected graphs, tournaments, digraphs, and $2-$structures.
We show that, under some axioms, a unique decomposition tree exists for umodules.
Polynomial-time algorithms are provided for:  non-trivial umodule test, maximal umodule computation, and decomposition tree computation when the tree exists.
Our results unify many known decomposition like modular and bi-join decomposition of graphs, and a new decomposition of tournaments.




\end{abstract}

\section{Introduction}
In graph theory modular decomposition is now a well-studied notion~\cite{Gal67,CHM81,MR84,ER90,EhrenfeuchtHR99}, as well as some of its generalisations~\cite{Cun73,HMC03,MontgolfierRaoICGT2005}.
As having been rediscovered in other fields, the notion also appears under various names, including intervals, externally related sets, autonomous sets, partitive sets, and clans.
Direct applications of modular decomposition include tractable constraint satisfaction problems,
computational biology,
graph clustering for network analysis, and graph drawing.

Besides, in the area of social networks, several vertex partitioning have been introduced in order to catch the idea of putting in the same part vertices acknowledging similar behaviour, in other words finding regularities~\cite{WhiteR83}.
Modular decomposition  provides such a partitioning, yet seemingly too restrictive for real life applications.
The concept of a role~\cite{EverettB91} on the other hand seems promising, however its computation unfortunately is $NP-$hard~\cite{FP03}.
As a natural consequence, there is need for the search of \emph{relaxed}, but \emph{tractable}, variations of the modular decomposition scheme.
A step following this direction has generalised graph modules to those of larger combinatorial structures, so-called homogeneous relations~\cite{BuixuanHLdM06,wg06,odsa2006}.
This paper follows the same research stream, and weakens the definition of module in order to further decompose. 
Fortunately we obtain a new tractable variation of modular decomposition, that we now introduce.

Modular decomposition is based on \emph{modules}, a vertex subset with no \emph{splitter}. In graphs, a splitter of a vertex subset is linked with some, but not all, vertices of this subset. We shall see how this definition can be extended to homogeneous relations.
The ``outside'' of a module constitutes therefore, for all vertices of the module, the same ordered partition.
For instance, all vertices of an undirected graph module have the same neighbourhood.
We here address unordered-modules, so-called \textit{umodules} for short: the outside of a umodule constitutes for all vertices of the umodule the same unordered partition.
For graph, the umodules are the \emph{bijoins} (see Fig.~\ref{fig_mod_umod_1} and Section~\ref{sec_appli}).
As there are clearly more umodules than modules, this allows deeper decomposition.
We shall see that this decomposition is tractable.


After comparing umodule to previous notions in the topic, we display its tractability by giving an $O(|X|^4\log|X|)$ time computation of the maximal umodules of a given homogeneous relation over a finite set $X$, and show how this can also be used as a non-trivial umodule existence test.
The structure of the family of umodules is then investigated under different scenarios.
We focus on a particular case, and provide a potent tractability theorem which makes use of the so-called \emph{Seidel-switching} graph operation~\cite{Seidel76}.
Fortunately enough, undirected graphs and tournaments fit into the latter formalism.
We then deepen the study and address total decomposability issues, namely when any ``large enough'' sub-structure is decomposable.
Surprisingly enough, this shows how our theory provides a very natural manner to obtain several results on \emph{round tournaments}
, including characterisation, recognition, and isomorphism testing (see e.g.~\cite{BJG01} for more detailed information), as well as further computational results, such as the \emph{feedback vertex set} computation.
\section{Umodule, an enlarged notion of module}
Let $X$ be a finite set.
The family of all subsets of $X$ is denoted by $\mP(X)$.
A \emph{reflectless} triple is $(x,y,z)\subseteq X^3$ with $x\ne y$ and $x\ne z$, which will be denoted by $(x|yz)$ instead of $(x,y,z)$ since the first element plays a particular role.
Let $H$ be a boolean relation over the reflectless triples of $X$.
Then, $H_x$ denotes the binary relation on $X\setminus\{x\}$ such that $H_x(y,z)\Leftrightarrow H(x|yz)$.
\begin{definition}[Homogeneous Relation and Module]\label{def1}\cite{BuixuanHLdM06,wg06,odsa2006}
$H$ is a \emph{homogeneous relation on $X$} if, for all $x\in X$, $H_x$ is an equivalence relation on $X\setminus\{x\}$
A subset $M\subseteq X$ is a \emph{module} of $H$ if $H(x|mm')$ for all $m,m' \in M$ and $x\in X\setminus M$.
\end{definition}

Equivalently, a homogeneous relation $H$ can be seen as a mapping from each $x\in X$ to a partition of $X\setminus\{x\}$, namely the equivalence classes of $H_x$.
This generalises graphs and \emph{2-structures}, where modular decomposition still applies under the different but equivalent name of \emph{clan decomposition}~\cite{EhrenfeuchtHR99,ER90}.
Roughly, a $2-$structure $G=(X,C)$ is a ground set $X$ and an edge colouration $C:X^2\rightarrow\mathbb{N}$ \cite{EhrenfeuchtHR99,ER90}.
Thus, a digraph is a $2-$structure using two colours, denoting the existing (when $C(x,y)=1$) and absent arcs (when $C(x,y)=0$).
There is no need of the concept of \emph{adjacency} nor \emph{neighbourhood} nor \emph{incidence} in a homogeneous relation!
But a homogeneous relation is canonically  derived from graphs and $2-$structures as follows.
\begin{definition}[Standard Homogeneous Relation]\label{def:standard}\cite{BuixuanHLdM06,wg06,odsa2006}
The \emph{standard homogeneous relation} $H(G)$ of a 2-structure $G=(X,C)$ is \hspace{2cm}
$H(G)(x|uv) \ \iff \ \  C(x,u)=C(x,v)\ \mathrm{and}\ C(u,x)=C(v,x).$



\end{definition}
\begin{proposition}\label{propo:standard}
Let $G$ be a graph, or a tournament, or an oriented graph, or a directed graph, or a $2-$structure.
The modules of  $H(G)$ exactly are the modules of $G$ in the usual sense (see definitions  in~\cite{Gal67,MR84,EhrenfeuchtHR99}).
\end{proposition}

We now introduce the central notion of this paper which, thanks to Proposition~\ref{prop_triviale} (below), can be seen at the same time as a proper generalisation of the classical modules/clans (in the sense of~\cite{Gal67,MR84,EhrenfeuchtHR99}), and a dual notion to the generalised modules (in the sense of~\cite{BuixuanHLdM06,wg06,odsa2006}).
\begin{definition}[Umodules]\label{def:umodul}
A subset $U$ of $X$ is a \emph{umodule} of $H$ if\\
\centerline{$\forall u,u' \in U, \ \ \forall x,x'\in X\setminus U, \ \ H(u|xx') \Longleftrightarrow H(u'|xx').$}
\end{definition}
\begin{figure}[t]
\begin{center}
	\subfigure[Undirected graphs]{\includegraphics[scale=0.35]{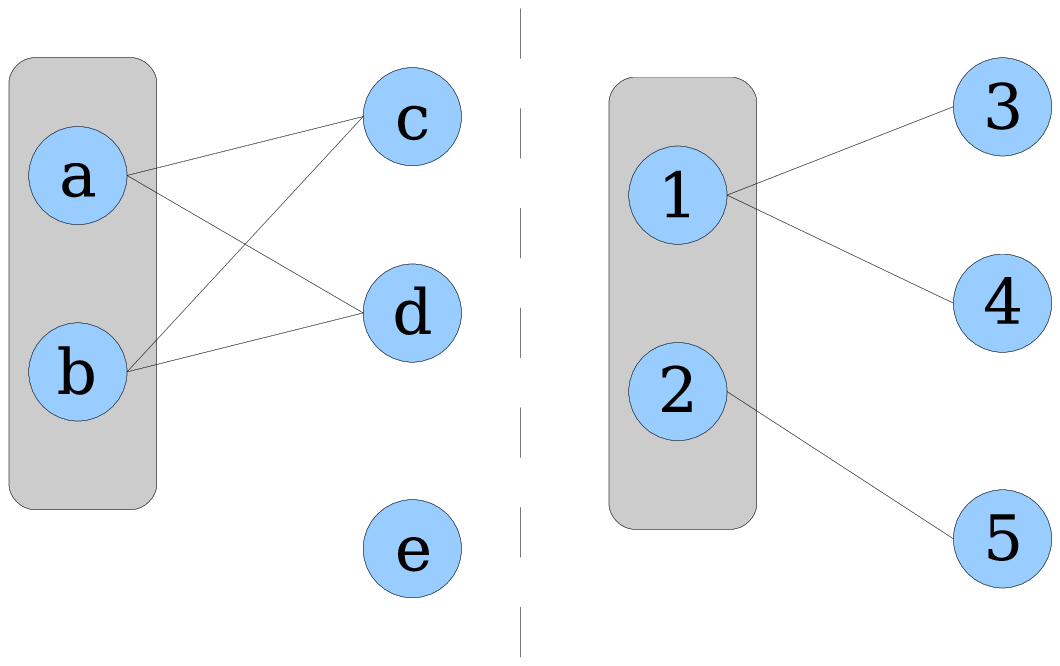}\label{fig_mod_umod_1}}
\hspace{1.5cm}
	\subfigure[Homogeneous relation]{\includegraphics[scale=0.9]{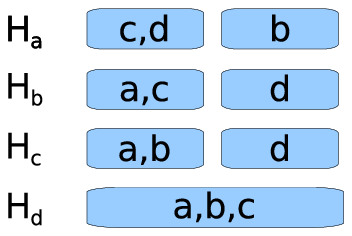}\label{fig_mod_umod_2}}
\caption{{\bf(a)} Modules and umodules in a graph: $\{a,b\}$ is a module and also a umodule, $\{1,2\}$ is a umodule but is not a module. {\bf(b)} A homogeneous relation with a module which is not a umodule.
$\{a,b\}$ is a module: they belong to the same equivalence class in both $H_c$ and $H_d$.
$\{a,b\}$ is not a umodule: $c$ and $d$ belong to the same class in $H_a$, and to different classes in $H_b$.}
\label{fig_mod_umod}
\end{center}
\vspace{-.5cm}
\end{figure}
Roughly, elements of a umodule come from the same ``school of thinking'':
 if one element of a umodule differentiates, resp. mixes together, some exterior elements, so does every element of the umodule (Fig.~\ref{fig_mod_umod}).
A umodule $U$ is \emph{trivial} if $|U|\le 1$ or if $|U|\ge|X|-1$.
The family of umodules of $H$ is denoted by $\mU_H$, and $\mU$ when no confusion occurs.
$H$ is \emph{umodular prime} if all umodules of $H$ are trivial.
The following  proposition links umodules to the $1$-intersecting families framework as defined in~\cite{HM85}.
The subsequent one tells how far umodules may generalise modules.
\begin{proposition}\label{propu}
For any two umodules $U,U'$ of a homogeneous relation $H$, if $U\cap U'\ne\emptyset$ then $U\cup U'$ is also a umodule of $H$.
\end{proposition}
\begin{proposition}\label{prop_triviale}
If $H$ is a standard homogeneous relation (see Definition~\ref{def:standard}), then any module of $H$ is a umodule of $H$.
If $H$ is an arbitrary homogeneous relation over a finite set $X$, then any module $M$ of $H$ is such that $X\setminus M$ is a umodule of $H$.
\end{proposition}

In case of graphs, a natural question arises~\cite{DC07}: for which graphs the notions of module and umodule coincide?
The following result, which can also be seen as a relaxed converse of Proposition~\ref{prop_triviale}, solves this problem.
As with modules, let the umodules of a graph refer to those of its standard homogeneous relation.
Notice here in a graph that the complementary of a umodule also is a umodule.
A \emph{threshold graph} is one that can be constructed from the single vertex by repeated additions of a single isolated or dominating vertex.
\begin{proposition}
\label{prop_threshold}
$G$ is a threshold graph if and only if in all induced subgraph of $G$, every umodule is either a module or the complementary of a module (or both).
\end{proposition}

Threshold graphs are known to be one of the smallest graph classes (see e.g.~\cite{BLS99}).
Therefore for most graphs umodules and modules differ, and Section~\ref{sec_appli} is devoted to the umodular graph decomposition.
However, before deepening decomposition issues, let us first display umodule tractability.
%
%
%
%
%
%
\section{Algorithmic Tractability for the general case}\label{sec_quotient}
As far as we are aware, 
 there is no evidence of a decomposition scheme for arbitrary umodules.
The first valuable objects to compute thus seem to be the maximal umodules with respect to some cut.
Using this, we also provide a polynomial time algorithm computing the \emph{strong} umodules (see definition afterwards).

\subsection{Maximal Umodules with respect to a cut}
Partitions will be ordered with respect to the usual partition lattice: $\mP=\{P_1,\dots,P_p\}$ is \emph{coarser} than $\mQ=\{Q_1,\dots,Q_q\}$, and $\mQ$ is \emph{thinner} than $\mP$, if every part $Q_i$ is contained in some $P_j$.
It is noted $\mQ\leq\mP$ and $\mQ<\mP$ if the partitions are different.
Let $S$ be a subset of $X$.
As the umodule family $\mU$ is closed under union of intersecting members (Proposition~\ref{propu}), the inclusionwise maximal umodules included in either $S$ or $X\setminus S$ form a partition of $X$, denoted by $MU(S)=MU(X\setminus S)$.
In other words, this is the coarsest partition of $X$ into umodules of $H$, which is thinner than $\{S, X\setminus S\}$.
Roughly, it gives an indication on how the umodules are structured with respect to $S$:
a umodule either is included in a umodule of $MU(S)$,
or properly intersects $S$,
or properly intersects $X\setminus S$,
or trivial.
\begin{definition}
Let $H$ be a homogeneous relation over $X$.
Let $C\subseteq X$.
The relation $R_C$ on $C$ is defined as:
\centerline{$ \forall x,y \in C,R_C(x,y) \ if \ \forall a,b \in (X\setminus C)  \ \  H(x|ab)  \Longleftrightarrow H(y|ab).$}
\end{definition}

This clearly is an equivalence relation on $C$.
Furthermore, $C$ is a umodule if and only if $R_C$ only has one equivalence class.
Let us define a \emph{refinement} operation, the main algorithmic tool for constructing $MU(S)$.
\begin{definition} 
Let $\mP$ be a partition of $X$ and $C$ a part of $\mP$. Let
$C_1,\ldots,C_k$ be the equivalence classes of $R_C$. $Refine(\mP,C)$ is the 
partition  obtained from  $\mP$, by replacing part $C$  by the  parts $C_1,\ldots,C_k$.
A partition $\mP$ is \emph{refinable} by $C$ if $Refine(\mP,C)\ne \mP$. 
$\mP$ is \emph{unrefinable} if for every part $C$ of $\mP$, we have $\mP=Refine(\mP,C)$.
\end{definition}
\begin{lemma}\label{lemf2}
Let $H$ be a homogeneous relation over $X$, $U$ a umodule of $H$, and $\mP$ a partition of $X$.
If $U$ is included in a part of $\mP$, then for any part $C$ of $\mP$, $U$ is included in a part of $Refine(\mP,C)$.
Moreover, a part $C$ of $\mP$ is a umodule if and only if $\mP$ is not refinable by $C$.
\end{lemma}

\begin{algorithm}[H]
\dontprintsemicolon
$\mP  \gets \{S,X\setminus S\}$ \;
\While{there exists an unmarked part $C$in $\mP$}{
	\lIf{$\mP=Refine(\mP,C)$}{
		mark $C$ \;
	}\lElse{
		$\mP\gets Refine(\mP,C)$ \;
	}
}
\caption{Refinement algorithm computing $MU(S)$ given homogeneous relation $H$ over $X$ and $S \subseteq X$}
\label{algo:Refinement}
\end{algorithm}

\bigskip

Correctness of Algorithm~\ref{algo:Refinement} follows from Lemma~\ref{lemf2} and the invariant: \textit{There is no umodule partition $\mQ$ such that $\mP < \mQ < \{S, X\setminus S\}$}.
So, starting from $\{S, X\setminus S\}$ the algorithm constructs a strictly decreasing chain of partitions of $X$ ending at $MU(S)$.
Let us see how to implement it efficiently.
\begin{lemma}\label{lemf3}
It is possible to compute  $Refine(\mP,C)$ in  $O(|X|^2)$ time. 
\end{lemma}
\begin{proof}
We first show how to test for $R_C(x,y)$.
Compute, for every element $x$ of $C$, a partition
$\mH(x,C)=\{P_x^1,\ldots,P_x^{k(x)}\}$ of $X\setminus C$. It is the
restriction of $H_x$ to $X\setminus C$, i.e $P_x^i=H_x^i \setminus C$. It
is easy to build in $O(|X|)$ time for each element of $C$.  Then we
have $R_C(x,y)$ if and only if $H(x,C)$ is exactly the same partition than
$H(y,C)$.  It can be tested in $O(|X|)$ time, but performing this for
each couple of elements of $C$ would lead to an $O(|X|^3)$ time
implementation of $Refine(\mP,C)$.
Let us instead consider  $\mH(x,C)$ as a $b$ bit vectors (with $b=|X\setminus
C|=O(|X|)$).  Looking for duplicates among these vectors can be
performed easily, by bucket sorting them on their first bit, then the
second, and so on. A scan of all vectors (i.e. of all elements of $C$)
compute the pairwise equal vectors, i.e the $R_C$ equivalent elements
of $C$. It is then easy to split $C$ and to update $\mP$, in  $O(|X|^2)$ time.
\qed
\end{proof}

The lemma above leads to an  $O(|X|^3)$ time implementation of Algorithm~\ref{algo:Refinement}. However,
\begin{theorem}\label{ThMU} 
For every $S \subseteq X$, $MU(S)$, the coarsest umodule partition  thinner than $\{S, X-S\}$  can be computed  in $O(|X|^2 \log|X|)$ time.
\end{theorem}
\begin{proof}
Using the well-known  Hopcroft's partition refinement rule it is possible to improve the above algorithm. The idea is to avoid at each step to consider the biggest part, see \cite{MR917035}.
Thus, to compute $MU(S)$ assuming that $|S| \leq |X-S|$, we first partition $X-A$ using the "neighbourhoods lists" of  all $a \in A$. If we assume a data structure which links each edge $ay$ to its opposite edge $ya$.
We can associate in the meantime to each element $a \in A$ a bitvector representing how $X-A$ sees $a$. These $|A|$ bitvectors of size $|X-A|$ can be sorted in $O(|X|.|X-A|) \in O(|X|^2)$.
Using Hopcroft's rule, a vertex $a$ can only be explored at most $O(\log|X|)$ time, which yields the announced complexity.
\qed
\end{proof}

\subsection{Strong Umodules: Maximal Umodules Computation and Primality Test}
A umodule is \emph{strong} if it overlap no other umodules, where two subsets overlap if none of the intersection and differences are empty.
As two strong umodules are either disjoint, or one contains another, they can be ordered by inclusion into a tree (see e.g. laminar families in~\cite{Schrijver03}). 
\begin{theorem}\label{thprim}
There exists an $O(|X|^4 \log|X|)$ algorithm to compute the inclusion tree of strong umodules.
\end{theorem}
\begin{proof}
Consider a non-trivial strong umodule $M$. For each pairwise distinct
$x,y\notin M$ (at least two of them exists since $M$ is not trivial),
$M$ is contained in exactly one set of $MU(\{x,y\})$. The intersection
of all these sets is exactly $M$. Indeed if it where $M'$ such that
$M\subsetneq M'$ then there would exist $x\in M'\setminus M$. For $y
\notin M$, $MU(\{x,y\})$ contains a umodule $M''$ smaller than $M'$
but containing $M$, a contradiction.
Then the algorithm is as follow. For every pair $\{x,y\}$ compute
$MU(\{x,y\})$ in $O(|X|^2 \log|X|)$ time (Theorem~\ref{ThMU}). That
gives a family of at most $|X|^3$ umodules. Add the trivial modules to
the family.  Greedily compute the intersection of overlapping
umodules of the family. It is possible in $O(|X|^4 \log|X|)$ time:
for each triple $(a,b,c)$ look for the umodules containing exactly two
of them, they overlap.  Then we have all strong umodules. We finally
just have to order them into a tree.
\end{proof}

This answers both maximal umodule computation and primality test since a non-trivial umodule exists if and only if a non-trivial strong umodule exists.

\section{Two Decomposition Scenarios}
Of course, the number of umodules may be as large as $2^{|X|}$. But we
shall now focus on certain umodule families having a compact
(polynomial-size) representation. Umodules of local congruence 2
relations, on the first hand, and self-complemented umodules families,
on the second hand, have such properties. They can be stored in
$O(|X|^2)$ and $O(|X|)$ space, respectively.

\subsection{Local Congruence and Crossing Families}\label{sectlc2} 
\begin{definition}[Local congruence]
Let $H$ be a homogeneous relation on $X$. For $x\in X$, the
\emph{congruence} of $x$ is the maximal number of elements that $x$
pairwise distinguishes.  In other words, it is the number of
equivalence classes of $H_x$. The \emph{local congruence} of $H$ is
the maximum congruence of the elements of $X$.
\end{definition}
\begin{Rq}
The standard homogeneous relation of an undirected graph  or a tournament has local congruence 2.
This value is 3 for an antisymmetric directed graph or a directed acyclic graph.
The value is 4 for digraphs.
\end{Rq}

When the local congruence of $H$ is $2$, so-call \emph{LC2 condition} for short, we obtain the following structural property on its umodule family.
\begin{definition}[Crossing family]
$\mathcal{F}\subseteq2^X$ is a crossing family if, for any $A, B \in \mathcal{F}$, that $A \cap B \neq \emptyset$ and $A \cup B \neq X$
implies $A \cap B \in \mathcal{F}$ and $A \cup B \in \mathcal{F}$ (see e.g.~\cite{Schrijver03} for further details).
\end{definition}

Crossing families commonly arise as the minimisers of a submodular function.
For instance, the minimum $s,t-$cuts of a network form a crossing family.
Gabow proved that a crossing  family admits a compact representation in $O(|X|^2)$ space using a tree representation~\cite{Gabow93}.
\begin{proposition}
The umodules of a homogeneous relation with local congruence $2$ form a crossing family, and can thus be stored in  $O(|X|^2)$ space.
\end{proposition}
%


\subsection{Self-complementarity and Bipartitive Families}\label{sec_self}
A consequence of previous proposition is that  standard homogeneous relations of graphs and tournaments have crossing umodules families.
But they have stronger properties, which we will use to show a linear-space structure coding the umodule family.
\begin{definition}[Four elements condition]\label{four}
$H$ fulfils the \emph{four elements
  condition} if \vspace{-.3cm}
\begin{eqnarray*}
\forall~m,m',x,x'\in X,~~
\left\{\begin{array}{l}
H(m|xx')\wedge  H(m'|xx')\wedge   H(x|mm') \Rightarrow H(x'|mm') \\
\neg H(m|xx') \wedge  \neg H(m'|xx')\wedge \neg H(x|mm') \Rightarrow \neg H(x'|mm')
\end{array}\right..
\end{eqnarray*}
\end{definition}
\begin{proposition}\label{prop_stand_4EC}
Standard homogeneous relations of undirected graphs and tournaments satisfy the four elements condition.
\end{proposition}

This is a light regularity condition, allowing to avoid examples similar to that of Fig.~\ref{fig_mod_umod_2}.
Surprisingly enough, it suffices to make the umodule family behave in a very tractable manner (Proposition~\ref{propo_4_self} and Corollary~\ref{coro_SC} below).
\begin{definition}[Self-complementary condition]
A family $\mF$ of subsets of $X$ is \emph{self-complemented} if for every subset $A$, $A\in\mF$ implies $X\setminus A\in\mF$.
\end{definition}
\begin{proposition}\label{propo_4_self}
If a homogeneous relation $H$ fulfils the four elements condition then the family $\mU$ of umodules of $H$ is self-complemented. 
\end{proposition}

The \emph{four elements condition} is quite convenient since it allows to shrink a umodule, hence apply the divide and conquer paradigm to solve optimisation problems.
However, as far as umodules are concerned, the \emph{self-complementary} relaxation is sufficient to describe a tree-decomposition theorem as can be seen in the following section.
Finally, notice that the converse of Proposition~\ref{propo_4_self} does not necessarily hold.
The characterisation of relations having a self-complemented umodule
family by a local axiom, such as the four elements condition, actually
appears to be more difficult.

\subsubsection{Tree Decomposition Theorem}
The following results on bipartitions can be found in \cite{Cun73} under the name of ``decomposition frame with the intersection and transitivity properties'',
in \cite{M03} under the name of ``bipartitive families'' (the formalism used in this paper),
and in \cite{HMC03} under the name of ``unrooted set families''.

We call $\{X_i^1,X_i^2\}$ a \emph{bipartition} of $X$ if $X_i^1\cup X_i^2=X$ and $X_i^1\cap X_i^2=\emptyset$.
Two bipartitions $\{X_i^1,X_i^2\}$ and $\{X_j^1,X_j^2\}$ \emph{overlap} if for all $a,b=1,2$ the four intersections $X_i^a\cap X_j^b$ are not empty.
A bipartition is \emph{trivial} if one of the two parts is of size $1$.
Let $\mB=\{\{X_i^1,X_i^2\}_{i\in 1,\ldots,k}\}$ be a family of $k$ bipartitions of $X$.
The \emph{strong} bipartitions of $\mB$ are those that do not overlap any other bipartition of $\mB$.
For instance, the trivial bipartitions of $\mB$ are strong bipartitions of $\mB$.
\begin{proposition}
If $\mB$ contains all trivial bipartitions of $X$, then there exists a unique tree $T(\mB)$ \vspace{-.3cm}
\begin{itemize}
\item with $|X|$ leaves, each leaf being labelled by an element of $X$.
\item such that each edge $e$ of  $T(\mB)$ correspond to a strong bipartition 
of $\mB$: the leaf labels of the two connected components of $T-e$ are exactly 
the two parts of a strong bipartition, and the converse also holds.
\end{itemize}
\end{proposition}

Let $N$ be a node of  $T(\mB)$ of degree $k$.
The labels of the leaves of the connected components of $T-N$ form a partition 
$X_1,\ldots,X_k$ of $X$. For $I\subseteq\{1,\ldots,k\}$ with $1<|I|<k$, the 
bipartition $B(I)$ is $\{\cup_{i\in I} X_i, X\setminus \cup_{i\in I}X_i\}$. \vspace{-.4cm}
\begin{definition}[Bipartitive Family]
A family of bipartitions is a \emph{bipartitive family} if it contains
all the trivial bipartitions and if, for two overlapping bipartitions
$\{X_i^1,X_i^2\}$ and $\{X_j^1,X_j^2\}$, the four bipartitions
$\{X_i^a\cup X_j^b, X\setminus (X_i^a\cup X_j^b)\}$ (for all
$a,b=1,2$) belong to $\mB$.
\end{definition}
\begin{theorem}\label{thbip}\cite{M03}
If $\mB$ is a bipartitive family, the nodes of $T(\mB)$ can be labelled \emph{complete}, \emph{circular}
or \emph{prime}, and the children of the \emph{circular} nodes can be ordered in such a way that: \vspace{-.3cm}
\begin{itemize}
\item If $N$ is a complete node, for any $I\subseteq \{1,\ldots,k\}$ such that   $1<|I|<k$, $B(I)\in\mB$.
\item If $N$ is a circular node, for any interval $I=[a,\ldots,b]$ of $\{1,\ldots,k\}$ such that   $1<|b-a|<k$,  $B(I)\in\mB$.
\item If $N$ is a prime node, for any element $I=\{a\}$ of $\{1,\ldots,k\}$ $B(I)\in\mB$.
\item There are no more bipartitions in $\mB$ than the ones described above.
\end{itemize}
\end{theorem}

For a bipartitive family $\mB$, the labelled tree $T(\mB)$ is an $O(|X|)$-sized representation of $\mB$, while the family can have up to $2^{|X|-1}-1$ bipartitions of $|X|$ elements each.
This allows to efficiently perform algorithmic operations on $\mB$.
Notice that any self-complemented subset family can be seen as a family of bipartitions.
\begin{proposition}\label{theorem_SC}
The members of a self-complemented umodule family form a bipartitive family.
\end{proposition}
\begin{corollary}[Umodular Decomposition Theorem]\label{coro_SC}
There is a unique $O(|X|)$-sized tree that gives a description of all possible umodules of a homogeneous relation $H$ fulfilling the self-complementary condition.
This tree is henceforth
called \emph{umodular decomposition tree}. Notice that it is an unrooted
tree, unlike the modular decomposition tree.
\end{corollary}
%
%
\subsubsection{Tree Decomposition Algorithm}
Let $H$ be a self-complemented homogeneous relation, $T(H)$ its umodular decomposition tree, and $U$ a nontrivial strong umodule (if any).
Let us examine some consequences of Theorem~\ref{thbip}. Notice
that two umodules overlap if and only if they are incident to the same node of
$T(H)$.  As $H$ is self-complemented the union of two overlapping
umodules is a umodule (Proposition~\ref{propu}) but also their
intersection.  The strong umodule $U$ is an edge in $T(H)$ incident
with two nodes $A$ and $B$. \vspace{-.3cm}
\begin{itemize}
\item If one of them,
say $A$, is labelled prime then for any $x,y\notin U$ such that the
least common ancestor of them in $T(H)$ is $A$, then $U\in
MU(\{x,y\})$.
\item If one of them, say $A$, is labelled circular then for any $x$ belonging 
to the subtree rooted in the successor of $U$ in the ordered circular list of 
$A$, and for any  $y$ belonging to the subtree rooted in the predecessor of 
$U$, then $U\in MU(\{x,y\})$.
\item  If one of them, say $A$, is labelled complete then the intersection, 
for all $x,y\notin U$ whose least common ancestor is $A$, the intersection of 
all parts of  $MU(\{x,y\})$ containing $U$ is exactly $U$.
\end{itemize}

Theorem~\ref{thprim} then can be used to compute the strong umodule inclusion tree.
After this, typing the nodes and ordering their sons according to the above definition is straightforward.
Hence,
\begin{theorem}
There exists an $O(|X|^4 \log(|X|))$ algorithm to compute the unique decomposition tree 
for a self complemented umodule family.
\end{theorem}
%
\section{Seidel-switching Theorem, a potent Tractability}
\begin{figure}[h]
	\begin{center}
		\subfigure[An example of a Seidel switch on an undirected graph]{\includegraphics[scale=.7]{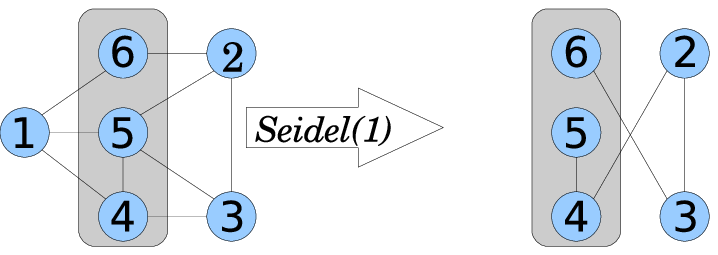} \label{fig:SeidelSwitchExemple}}
\hspace{0.5cm}
		\subfigure[1. A bi-join (i.e. umodule) in an undirected graph, 2. a umodule in a tournament]{\includegraphics[scale=.7]{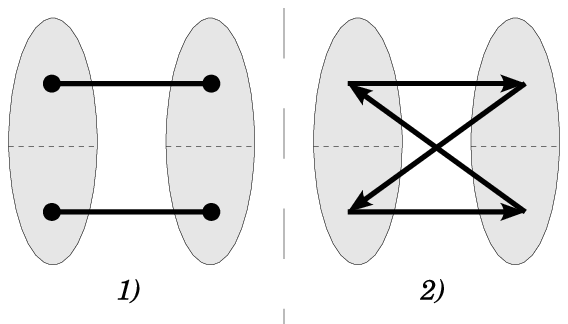} \label{figbij}}
\vspace{-0.3cm}
		\caption{(a) Seidel switching, (b) umodules on undirected graphs}
	\end{center}
\end{figure}

Standard homogeneous relations of graphs and tournaments are of local congruence 2, and their umodule families are self-complemented.
Firstly this means we can either decompose those families using the crossing families decomposition or using the bipartitive decomposition.
Moreover, relations that satisfy both the self-complementary and \emph{LC2} properties seem to own stronger potential.
In particular, let us show a nice local transformation from the umodules of such a relation  to the modules of another relation.
This operation was first introduced in J. Seidel in \cite{Seidel76} on undirected graphs.
It was later studied by several authors interested in some computational aspects \cite{ColbournC80,KratochvilNZ92} and structural properties 
\cite{Hayward96,Hertz99} and recently in \cite{MontgolfierRaoICGT2005}.
The operation is referred to as \emph{Seidel switch} in \cite{Hertz99}, and we will adopt this terminology.
We generalise it to homogeneous relations but take a restricted case of switch, with the slight difference  
that we remove from the transformation an element (see Fig.~\ref{fig:SeidelSwitchExemple}).
For convenience, if $H$ is a homogeneous relation on $X$ and $s\in X$,
we also refer to the equivalence classes of $H_s$ as $H_s^1,\dots,H_s^k$.
\begin{definition}[Seidel switch]
\label{def:SeidelSwitch}
Let $H$ be a homogeneous relation of local congruence $2$ on $X$,
and $s$ an element of $X$.
The \emph{Seidel switch} at $s$ transforms $H$ into the homogeneous relation $H(s)$ on $X\setminus\{s\}$ defined as follows.
$$\forall x\in X\setminus\{s\}, H(s)^1_x=(H_x^1\Delta H_s^j)\setminus\{s\} \ and \  H(s)^2_x=(H_x^2\Delta H_s^j)\setminus\{s\}$$
with $j$ such that $x \notin H_{s}^{j}$. 
where $A\Delta B$ denotes the symmetric difference of $A$ and $B$.
\end{definition}
%
%
\begin{theorem}[Seidel-switching Theorem]\label{thmain}
Let $H$ be a homogeneous relation of local congruence $2$ on $X$ such that $\mU_H$ is self-complemented.
Let $s$ be a member of $X$, and $U\subseteq X$ a subset containing $s$.
Then, $U$ is a umodule of $H$ if and only if $M=X\setminus U$ is a module of the Seidel switch $H(s)$.
\end{theorem}
\begin{corollary}\label{corX}
The umodular decomposition tree of a self-complemented homogeneous relation of local congruence $2$ on $X$ can be computed in $O(|X|^2)$ time.
\end{corollary}
\begin{proof}
Using a Seidel switch on any element will result in a relation having the so-called \emph{modular quotient} property~\cite{BuixuanHLdM06}: every module of the relation also is a umodule.
Then, the $O(|X|^2)$-time modular decomposition algorithm for modular quotient relations depicted in~\cite{BuixuanHLdM06}. As two complemented strong umodules $M, X\setminus M$ of $H$, for $s\notin M$, correspond to a strong module $M$ of $H(s)$, then 
the strong umodules of $H$ can be found trivially from the strong modules of $H(s)$. Typing and ordering their sons is then easy. 
\end{proof}

Notice that the modular
 decomposition tree of $H$ can be trivial, 
while the one of its Seidel switch at $s$ may be not.
Besides, there is no real need to type and order the sons of a node, as
so-called \emph{linear} nodes of the modular decomposition tree give
circular nodes of the umodular decomposition tree with the same
ordering of their sons, complete nodes of $H(s)$ give complete nodes
of $H$ and prime nodes of $H(s)$ give prime nodes of $H$. The
correspondence is straightforward but modular decomposition of
homogeneous relations will not be discussed here, the reader should
refer to~\cite{BuixuanHLdM06}.

\section{Umodular Decomposition of Graphs and Tournaments}\label{sec_appli}
Let us now apply umodular decomposition to two well-known
combinatorial objects: undirected graphs and tournaments. In this
section we always implicitly refer to their standard homogeneous
relations, for instance ``the umodules of the graph $G$'' stands for
``the umodules of the standard homogeneous relation $H(G)$ of the graph
$G$'' and so on. And ``graph'' stands for ``undirected graph''. As we
have seen, graphs and tournaments fulfil the four elements
conditions, are of local congruence two, and their umodule family is
self-complemented.

\subsection{Bijoin decomposition}
Let us call \emph{bijoin} a umodule of a graph or of a tournament.
From definition, one can see what bijoins are (Fig.\ref{figbij}).
In a graph, $B$ is a bijoin if $X\setminus B$ can be partitioned in two sets $C$ and $D$ such that for each $x\in B$, either $N(x)\cap C = \emptyset$ and $D\subseteq N(x)$, or  $N(x)\cap D = \emptyset$ and $C\subseteq N(x)$.
For a tournament, same definition with  $C\subseteq N^+(x)$ and  $D\subseteq N^-(x)$, or  $D\subseteq N^+(x)$ and  $C\subseteq N^-(x)$.

Bijoins of graphs where studied in \cite{MontgolfierRaoICGT2005}
as a new graph decomposition, generalising modular decomposition.  The
Seidel switch was used to derive most of the properties claimed,
especially a decomposition tree (with no \emph{circular} nodes), a
linear-time decomposition algorithm, a characterisation of the two
kinds of \emph{complete} nodes, and characterisation of \emph{totally
decomposable} graphs (see below).

Bijoins of tournaments form a new decomposition. The first important property is:
\begin{proposition}
The umodular (bijoin) decomposition tree computation time of a tournament is $O(|X|^2)$.
\end{proposition}

The tree exists thanks to Corollary~\ref{coro_SC}, since the bijoins form a self-complemented $LC2$ family. The computation algorithm is from  Corollary~\ref{corX}.
 \begin{proposition}
 The umodular (bijoin) decomposition tree of a tournament has no \emph{complete node}. And there exists a circular ordering of the vertices of the tournament
 such that every umodule of the tournament is a factor (interval) of this
 circular ordering.
 \end{proposition}

The first assumption can be checked by reader: it is impossible to
build tournaments with more than four elements such that every vertex
subset is a bijoin. The second is a consequence of the first, and of
definitions in Theorem~\ref{thbip}.  As a consequence, there are
$O(|X|^2)$ bijoins in a tournament (the exponential growth of a
bipartitive family comes from \emph{complete} nodes).


\subsection{Total Decomposability}
Given a graph decomposition scheme, is often worth to consider the totally decomposable graphs with respect to that scheme, namely the graphs in which every "large enough" subgraph admits a non trivial decomposition.
In general this leads to the definition of very interesting class of graphs, such as cographs with modular decomposition or distance hereditary graphs with split decomposition.
Let us now see how the graphs and tournaments totally decomposable with respect to bijoin decomposition behave.

\begin{theorem}\cite{MontgolfierRaoICGT2005}
The totally decomposable graph with respect to bijoin decomposition are the ($C_5$,bull,gem,co-gem)-free
graphs, and also exactly the graphs that can be obtained from a single
vertex by a sequence of (twin,antitwin)-extensions.
\end{theorem}

\begin{figure}[h!]
 	\begin{center}
 		\includegraphics[scale=.55]{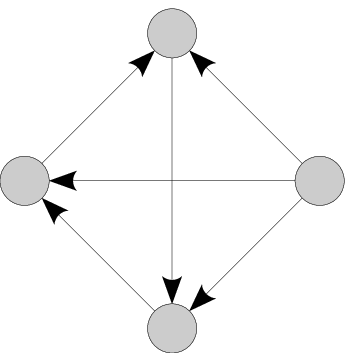}
 		\hspace{2cm}
 		\includegraphics[scale=.55]{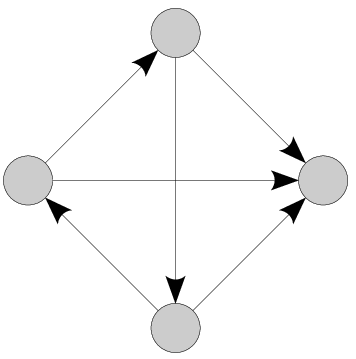}
 	\end{center}
 	\caption{ Forbidden 
 subgraphs of a totally decomposable tournament with respect to umodular decomposition: the in-diamond(left) and the out-diamond(right). }
 \label{fig:PrimeConfTour}
 \end{figure}

\begin{definition}
A \emph{diamond} is one of the induced 
subgraph described in Figure~\ref{fig:PrimeConfTour}.
A tournament $T$ is \emph{locally transitive}
if
for each vertex $x \in V(T)$, $T_{[N^{+}{(x)}]}$ and $T_{[N^{-}{(x)}]}$ are transitive 
 tournaments.
Two vertices $x$ and $y$ of a tournaments are \emph{twins} if $N^+(x)\setminus\{y\} = N^+(y)\setminus\{x\}$ and \emph{antitwins} if $N^+(x)\setminus\{y\} = N^-(y)\setminus\{x\}$.
An \emph{extension} of a vertex $x$ of $T$ by a twin (resp. antitwin) $y$ consists in adding a new vertex $y$ to $T$ and making $y$ twin (resp. antitwin) of $x$.
 \end{definition}

\begin{theorem}
Let $T$ be a tournament. The following propositions are equivalent:
\begin{enumerate}
\item $T$ is diamond-free (no induced subgraph is a diamond)
\item $T$ is locally transitive
\item  $T$ is totally decomposable with respect to bijoin decomposition
\item $T$ can be obtained from a  single
vertex by a sequence of (twin,antitwin)-extensions.
\end{enumerate}
\end{theorem}
\begin{proof}
As the in- and out-diamond are prime with respect to umodular decomposition, and total decomposability is an hereditary property, Point 3 implies Point 1. 
Let us sketch the proof that Point 2 implies Point 3. If $T$ is locally transitive then a Seidel switch of $T$ at any vertex $s$ produces a transitive tournament $T(s)$. Every subgraph of a transitive tournament contains a module. So, according to Theorem~\ref{thmain}, every subgraph of $T$ contains a umodule: $T$ is totally decomposable.
Besides, the equivalence between Point 3 and Point 4 comes from the fact that, if $T$ is totally decomposable, then it contains a umodule of two vertices. Such umodules are made either with two twins or with two antitwins.
Equivalence between 1 and 2 can be found in \cite{BJG01}.
\end{proof}

It is not hard to check that, as the umodular decomposition tree of a totally decomposable tournament may have no \emph{prime} node, and since two \emph{circular} node may not be adjacent, then the umodular decomposition tree  of a totally decomposable tournament has only a single \emph{circular} node. The ordering of the vertices along this node is known as \emph{circular ordering}. This ordering is such that, for each vertex $x$, the vertices of $N^+(x)$ follow consecutively; and  so do vertices from $N^-(x)$.
This, combined to the above theorem, could be seen as a sketched proof of the characterisation of \emph{round tournaments} by local transitivity (see e.g.~\cite{BJG01} for further information).

In the extended version \cite{BuixuanHLdM07a}, we present an  $O(n^2)$ recognition algorithm, 
 making an intensive use of this ordering property, and computing this ordering. It allows us to solve the 
isomorphism problem for the class of such tournament in  $O(n^2)$ time, like in \cite{Clarou96}.
We also propose the first linear-time algorithm for the feedback
vertex set problems (NP-complete for tournaments).  The 
basic idea  is to find a vertex of highest outgoing degree, and output
the tournament composed of this vertex and its outgoing neighbourhood.

\section{Extensions and further developments}
We have presented the umodules and homogeneous relations focusing on
graph theory field. But umodules may be found in many other objects.
For instance, if we take a commutative ring and define\\
\centerline{$H_\times(x|yz) \ \iff \ xy = xz,$}
then  the principal ideals of the ring are umodules.
In this paper we study umodular decomposition applied to graphs, when the local congruence is 2, the next challenge is now
 to understand  umodular decomposition of directed graphs or directed acyclic graphs, starting with the self-complemented case first.

Our  computation of strong umodules is polynomial, but its asymptotic
complexity of $O(|X|^4. \log(|X|))$ can surely be reduced, especially when applied to particular 
combinatorial objects.

We have noticed here the great importance of  the Seidel switch operation, and following the notion of vertex minor as defined in \cite{Oum05a},
let us called $H$ a \emph{Seidel minor} of a graph $G$, if $H$ can be obtained from
$G$ by the two following operations: 
\begin{itemize} \vspace{-.1cm}
\item
delete a vertex, 
\item choose a vertex and do a Seidel switch on this vertex
\end{itemize} \vspace{-.1cm}
It could be of interest to study  such Seidel minors.


%
%
%
%
%
{
\small
\bibliography{ShortUmodule}
\bibliographystyle{plain}
}
\end{document}